\documentclass{elsart}
\usepackage{graphicx}
\usepackage{amssymb}

\begin{document}
\begin{frontmatter}
\title{The parabolic approximation of channeling and diffraction in bent crystals}

\author{Gennady V. Kovalev}
\author{}
\address{School of Mathematics \\University of Minnesota, Minneapolis, MN 55455,USA}

\begin{abstract}
The parabolic approximation is developed for high energy charged particles 
scattering in a bent crystal with variable curvature. The general form of parabolic equation is received for atomic chains located along coordinate axis of orthogonal curvilinear coordinate system. 
\end{abstract}

\end{frontmatter}

\section{\label{sec:level1}Introduction}
The parabolic equation was introduced by Leontovich and Fock~\cite{leontovich46} to describe the scattering of the wave in a cone centered on a direction of wave propagation. Lervig, Lindhard and Nielsen~\cite{lln67} tried to use the similar time dependent Schr¨odinger equation for quantum treatment the directional effects of energetic charged particles in the crystals. The work~\cite{lln67} indicated that an asymptotic expansion of the wave equation with large parameter $p*R \gg 1$ ($p$ - the incident momentum of particle, $R$ - the atomic screen radius) should not be done in the direction of $p$, but in the direction of crystal axes (planes), which is slightly different from the direction of  the particle. The work~\cite{kov_1_85} shown that accurate calculations of potential scattering in the field of a straight atomic chain can be done as a double asymptotic expansion in terms of two large parameters of the scattering: the length of the atomic chain $L_x/R \gg 1$ and the component of the incident momentum of particle along the chain $p_x*R \gg 1$. It is important to note~\cite{kov_1_85,kov_2_85} that this double expansion gives the parabolic equation as well as the model of continuum potential which is central idea of channeling effect.
The expansion along the axis of symmetry is the major difference between original parabolic equation~\cite{leontovich46} and its applications for the scattering in the crystals. The parabolic equation method is also more general than sudden collision approximation~\cite{molier47} or so called eikonal-type approximations ~\cite{ll3,schiff56}, and semi-classical corrections to the eikonal scattering amplitude can be derived from it. 
  
  In this report we receive the parabolic equations using the symmetry of the crystals  described by orthogonal curvilinear coordinate system. On local level, such systems looks as Cartesian coordinate systems. However, there is a difference between the standard parabolic equation~\cite{leontovich46} and the parabolic equations in curvilinear coordinate systems considered here.  

\section{\label{sec:02}Parabolic equations in cylindrical symmetry}
\begin{figure}
	\centering
		\includegraphics{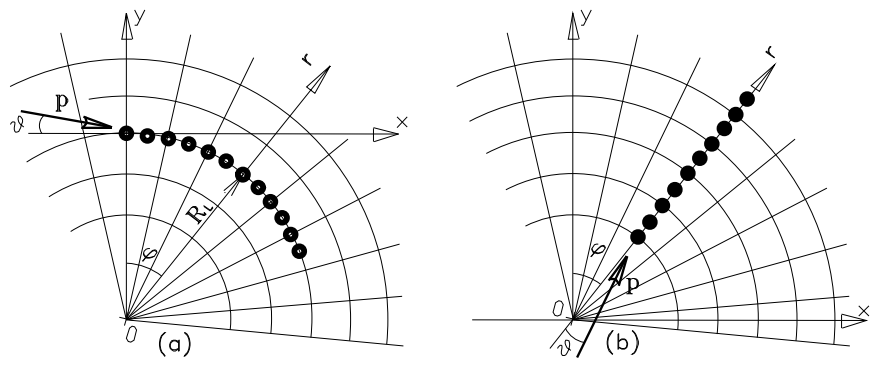}
	\caption{Atomic chain along (a)the circumference and (b) radius of circle }
	\label{parab1a}
\end{figure}

The parabolic approximation allows to construct a family of solutions to the Schr¨odinger wave equation which are close to the plane wave along some direction. If the potential of scattering $U$ has a spherical symmetry, the only possible direction of expansion is the direction of incidence $\vec{p}$ of the particle. Denoting this direction as $x$, the wave function $\Psi$	satisfying to the stationary Schr¨odinger equation  
\begin{eqnarray}
(\Delta+p^2-V)\Psi=0, 
\label{schrod}	
\end{eqnarray}
and can be presented as
\begin{eqnarray}
\Psi=exp(i p x)\Phi(\vec{r}),
\label{wfun_1}	
\end{eqnarray}
where $V=2MU$,  $M$ is the mass of particle ($\hbar=c=1$), $\Phi(\vec{r}$ is very slowly varying in the direction $x$ function in comparison to the $exp(i p x)$. Substitution of Eq. ~(\ref{wfun_1}) into ~(\ref{schrod}) and neglecting of the term $\partial^2 \Phi/\partial x^2$ yields a well known parabolic equation~\cite{leontovich46,fock65}
\begin{eqnarray}
2 i p \;\partial \Phi/\partial x +(\Delta_\bot -V)\Phi=0,
\label{par_1}	
\end{eqnarray}
with transverse Laplacian $\Delta_\bot = \partial^2 /\partial y^2+\partial^2 /\partial z^2$. When the scattering is studied on potential centers which constitute a straight atomic chain (or plane), and the momentum of particle is not parallel to the axis $x$ of chain, $\vec{p}\vec{r}=p_x x+\vec{p_\bot}\vec{r}_\bot$ , the the wave function $\Psi$ should reflect the translation symmetry along the direction of the chain
\begin{eqnarray}
\Psi=exp(i p_x x)\Phi(\vec{r}).
\label{wfun_2}	
\end{eqnarray}
Then the parabolic equation becomes
\begin{eqnarray}
2 i p_x \;\partial \Phi/\partial x +(\Delta_\bot+p^2-p_x^2 -V)\Phi=0.
\label{par_2}	
\end{eqnarray}
Now we turn to the parabolic equation for bent atomic chain. First, consider a circlular bent with a constant curvature $k = 1/R_l$ and parametric presentation of a chain along this circle $x=R_l \sin(\phi), y=R_l (\cos(\phi)-1), z=z$ (see Fig.~\ref{parab1a}(a)).
If we take a cylindrical coordinate system ($\rho,\phi,z$), then there are
the conjugate momentums of free particle ($p_{\rho}, p_{\phi}, p_z$), such that
\begin{eqnarray}
\vec{p} \vec{r} = p_{\rho} \rho + p_{\phi} \phi + p_z z.	
\end{eqnarray}
Based on this presentation of phase, we can derive three different parabolic equations, each one for the propagation of the particle along one  curvilinear coordinate when two others are constants. Parabolic equation along the axis $z$ lookes the same as \ref{par_2}, so we consider two cases in Fig.~\ref{parab1a}. For the motion Fig.~\ref{parab1a}(a) the wave function can be in the form
\begin{eqnarray}
\Psi=exp(i p_{\phi} \phi ) \Phi(\rho,\phi,z).
\label{wfun_3}	
\end{eqnarray}
Using Laplacian in cylindrical coordinates   
\begin{eqnarray}
\Delta =\frac{1}{\rho^2} \frac{\partial^2}{\partial \phi^2} + \frac{\partial^2}{\partial \rho^2} +\frac{1}{\rho} \frac{\partial}{\partial \rho} + \frac{\partial^2}{\partial z^2},
\end{eqnarray}
we obtaine the eq.
\begin{eqnarray}
\frac{2 i p_\phi}{\rho^2} \frac{\partial \Phi}{\partial \phi}+ \frac{1}{\rho^2}  \frac{\partial^2\Phi}{\partial \phi^2}+(\Delta_\bot+p^2-\frac{p_\phi^2}{\rho^2} -V)\Phi=0
\end{eqnarray}
with transverse 2D Laplacian $\Delta_\bot = \partial^2 /\partial \rho^2+1/\rho \partial /\partial \rho + \partial^2 /\partial z^2$. The next step is to neglect the small term $1/\rho^2  \partial^2\Phi /\partial \phi^2$ in range of high energies, and we have the final result
\begin{eqnarray}
\frac{2 i p_\phi}{\rho^2} \frac{\partial \Phi}{\partial \phi}+ (\Delta_\bot+p^2-\frac{p_\phi^2}{\rho^2} -V)\Phi=0.
\label{par_3}
\end{eqnarray}
In the second case (Fig.~\ref{parab1a}(b)), the wave function can be written as
\begin{eqnarray}
\Psi=exp(i p_{\rho} \rho ) \Phi(\rho,\phi,z),
\label{wfun_4}	
\end{eqnarray}
and doing the similar procedure, one can get the parabolic equation for the radial motion
\begin{eqnarray}
2 i p_\rho \frac{\partial \Phi}{\partial \rho}+\frac{1}{\rho}\frac{\partial \Phi}{\partial \rho}+ (\Delta_\bot+p^2-p_{\rho}^2 + i\frac{p_\rho}{\rho} -V)\Phi=0.
\label{par_4}
\end{eqnarray}
However, using the assumption $p_\rho \rho \gg 1$ the terms $1/\rho\partial \Phi/\partial \rho$ and $ip_\rho/\rho$ can be disregarded in Eq.(\ref{par_4}) and parabolic equation becomes the same as  Eq.(\ref{par_2}) for the straight chain. 
\section{\label{sec:03}Parabolic equations in orthogonal curvilinear coordinate system}

Now we derive the general form the parabolic equation. Assume an atomic chain is located along the curvilinear axis $q_1$ of some orthogonal curvilinear coordinate system $(q_1,q_2,q_3)$. The Laplacian in this coordinate system can be written in the form
\begin{eqnarray}
\Delta=\frac{1}{h_1 h_2 h_3}\left(\frac{\partial}{\partial q_1}(\frac{h_2 h_3}{h_1}\frac{\partial}{\partial q_1})+\frac{\partial}{\partial q_2}(\frac{h_1 h_3}{h_2}\frac{\partial}{\partial q_2})+\frac{\partial}{\partial q_3}(\frac{h_1 h_2}{h_3}\frac{\partial}{\partial q_3})\right)
\label{Laplace}	
\end{eqnarray}
where $h_i = \sqrt{g_{ii}}$ is Lame coefficients; $g_{ij}$ is metric tensor of the space, $g_{ij}=0$ for $i \neq j$. For example, in the cylindrical  coordinate system considered in Sec.\ref{sec:02}  $h_{1}=1, h_{2}=\rho, h_{3}=1$ and , in the elliptic  coordinate system $x=R_l \cos(q_2) \cosh(q_1),\: y=R_l \sin(q_2) \cosh(q_1),\: z=q_3$ Lame coefficients are $h_{1}=h_{2}=R_l \sqrt{\cosh(q_1)^2-cos(q_1)^2}, \:h_{3}=1$ .
The wave function may be presented as (obviously, due to symmetry there is no difference what coordinate $q_i$ can be chosen)
\begin{eqnarray}
\Psi=exp(i p_{q_1} q_1)\Phi(q_1,q_2,q_3).
\label{wfun_5}	
\end{eqnarray}
The possibility of such presentation is based on relation between direct $(q_1,q_2,q_3)$ and dual$(p_{q_1},p_{q_2},p_{q_3})$ spaces
\begin{eqnarray}
	\vec{p}\vec{r}=p_{q_1}q_1+p_{q_2}q_2+p_{q_3}q_3.
\end{eqnarray}
 
Acting in the same manner as in Sec.~\ref{sec:02} , we can derive three different parabolic equations, each one for propagation of the particle along one  curvilinear coordinate when two others are constants. We proceed with the wave function (\ref{wfun_5}).
Substituting (\ref{wfun_5}) in wave equation (\ref{schrod}) with Laplacian (\ref{Laplace}), the accurate equation for reduced wave function $\Phi$ can be written as

\begin{eqnarray}
\frac{2 i p_{q_1}}{h_1^2} \frac{\partial \Phi}{\partial q_1}+\frac{1}{h_1 h_2 h_3} \frac{\partial}{\partial q_1}(\frac{h_2 h_3}{h_1})\frac{\partial \Phi}{\partial q_1}+\frac{1}{h_1^2}\frac{\partial^2 \Phi}{\partial q_1^2}+ \nonumber\\
+(\Delta_\bot+p^2-\frac{p_{q_1}^2}{h_1^2}+i\frac{p_{q_1}}{h_1 h_2 h_3}\frac{\partial}{\partial q_1}(\frac{h_2 h_3}{h_1}) -V)\Phi=0.
\label{par_5a}	
\end{eqnarray}
where the tranverse Laplacian is
\begin{eqnarray}
\Delta_\bot =\frac{1}{h_1 h_2 h_3}\left(\frac{\partial}{\partial q_2}(\frac{h_1 h_3}{h_2}\frac{\partial}{\partial q_2})+\frac{\partial}{\partial q_3}(\frac{h_1 h_2}{h_3}\frac{\partial}{\partial q_3})\right).
\end{eqnarray}
Now since the second derivative $\frac{1}{h_1^2}\frac{\partial^2 \Phi}{\partial q_1^2}$ in (\ref{par_5a}) has relatively small value, the parabolic equation in orthogonal curvilinear coordinates becomes
\begin{eqnarray}
\frac{2 i p_{q_1}}{h_1^2} \frac{\partial \Phi}{\partial q_1}+\frac{1}{h_1 h_2 h_3} \frac{\partial}{\partial q_1}(\frac{h_2 h_3}{h_1})\frac{\partial \Phi}{\partial q_1}+ \nonumber\\
+(\Delta_\bot+p^2-\frac{p_{q_1}^2}{h_1^2}+i\frac{p_{q_1}}{h_1 h_2 h_3}\frac{\partial}{\partial q_1}(\frac{h_2 h_3}{h_1}) -V)\Phi=0.
\label{par_5}	
\end{eqnarray}
This is the most general form of parabolic equation, and all results received in Sec.\ref{sec:02} can easily be deduced from it. Indeed, in Cartesian coordinates $q_{1}=x, q_{2}=y, q_{3}=z$ and $h_{1}=h_{2}=h_{3}=1$, now from (\ref{par_5}) it follows the Eq.(\ref{par_2}). In cylindrical coordinates $q_{1}=\rho, q_{2}=\phi, q_{3}=z$ and $h_{1}=1, h_{2}=\rho, h_{3}=1$.  From (\ref{par_5}) it follows the Eq.(\ref{par_4}). 
Parabolic equations along the other axes $q_2, q_3$ look exactly the same as (\ref{par_5}) with cyclic substitution of indices. For example, the parabolic equation along the axis $q_2$ will have the form
 \begin{eqnarray}
\frac{2 i p_{q_2}}{h_2^2} \frac{\partial \Phi}{\partial q_2}+\frac{1}{h_1 h_2 h_3} \frac{\partial}{\partial q_2}(\frac{h_1 h_3}{h_2})\frac{\partial \Phi}{\partial q_2}+ \nonumber\\
+(\Delta_\bot+p^2-\frac{p_{q_2}^2}{h_2^2}+i\frac{p_{q_2}}{h_1 h_2 h_3}\frac{\partial}{\partial q_2}(\frac{h_1 h_3}{h_2}) -V)\Phi=0.
\label{par_6}	
\end{eqnarray}
Now it is straightforward to see that in cylindrical coordinates Eq.(\ref{par_6}) has the form of the Eq.(\ref{par_3}).

\bibliographystyle{elsart-num}
\bibliography{../../../Focusing_and_Channeling_in_Crystals/chan02}

\end{document}